\documentclass[reprint,superscriptaddress,showpacs,preprintnumbers,showkeys,amsmath,amssymb,aps,prb,floatfix]{revtex4-1}
\usepackage{graphicx}
\usepackage{bm}
\usepackage[colorlinks=true, citecolor=blue, urlcolor=blue, linkcolor=blue]{hyperref}
\usepackage[normalem]{ulem}
\usepackage{soul}

\newcommand{\CCOC}{Ca$_2$CuO$_2$Cl$_2$}
\newcommand{\SCOC}{Sr$_2$CuO$_2$Cl$_2$}
\newcommand{\NaCCOC}{Ca$_{2-x}$Na$_x$CuO$_2$Cl$_2$}
\newcommand{\VCCOC}{Ca$_{2-x}$CuO$_2$Cl$_2$}
\newcommand{\LSCO}{La$_{2-x}$Sr$_x$CuO$_4$}
\newcommand{\LCO}{La$_{2}$CuO$_4$}

\begin{document}

\title{Resonant inelastic X-ray scattering study of spin-wave excitations in the cuprate parent compound \texorpdfstring{Ca$_{2}$CuO$_2$Cl$_2$}{CCOC}}

\author{B. W. Lebert}
\affiliation{IMPMC-Sorbonne Universit\'es, Universit\'e Pierre et Marie Curie, CNRS, IRD, MNHN 4, place Jussieu, 75252 Paris, France}
\affiliation{Synchrotron SOLEIL, L'Orme des Merisiers, Saint-Aubin, 91192 Gif-sur-Yvette Cedex, France}

\author{M. P. M. Dean} 
\affiliation{Department of Condensed Matter Physics and Materials Science, Brookhaven National Laboratory, Upton, USA}

\author{A. Nicolaou} 
\affiliation{Synchrotron SOLEIL, L'Orme des Merisiers, Saint-Aubin, 91192 Gif-sur-Yvette Cedex, France}

\author{J. Pelliciari} 
\author{M. Dantz} 
\author{T. Schmitt}
\affiliation{Swiss Light Source, Paul Scherrer Institut, CH-5232 Villigen PSI, Switzerland}

\author{R. Yu}
\author{M. Azuma}
\affiliation{Materials and Structures Laboratory, Tokyo Institute of Technology, 4259 Nagatsuta, Midori, Yokohama 226-8503, Japan}

\author{J-P. Castellan}
\affiliation{Laboratoire L\'{e}on Brillouin (CEA-CNRS), CEA-Saclay, F-91911 Gif-sur-Yvette, France}
\affiliation{Institute for Solid State Physics, Karlsruhe Institute of Technology, D-76021 Karlsruhe, Germany}

\author{H. Miao} 
\affiliation{Department of Condensed Matter Physics and Materials Science, Brookhaven National Laboratory, Upton, USA}

\author{A. Gauzzi}
\author{B. Baptiste}

\author{M. d'Astuto}
\email{matteo.dastuto@impmc.upmc.fr}
\affiliation{IMPMC-Sorbonne Universit\'es, Universit\'e Pierre et Marie Curie, CNRS, IRD, MNHN 4, place Jussieu, 75252 Paris, France}

\date{\today}

\begin{abstract}
By means of resonant inelastic x-ray scattering at the Cu L$_3$ edge, we measured the spin wave dispersion along $\langle$100$\rangle$ and $\langle$110$\rangle$ in the undoped cuprate \CCOC. The data yields a reliable estimate of the superexchange parameter $J$ = 135 $\pm$ 4~meV using a classical spin-1/2 2D Heisenberg model with nearest-neighbor interactions and including quantum fluctuations. Including further exchange interactions increases the estimate to $J$ = 141~meV. The 40~meV dispersion between the magnetic Brillouin zone boundary points (1/2,\,0) and (1/4,\,1/4) indicates that next-nearest neighbor interactions in this compound are intermediate between the values found in \LCO{} and \SCOC{}. Owing to the low-$Z$ elements composing \CCOC, the present results may enable a reliable comparison with the predictions of quantum many-body calculations, which would improve our understanding of the role of magnetic excitations and of electronic correlations in cuprates.
\end{abstract}



\maketitle
\section{\label{intro}Introduction}
Magnetic excitations have been intensively studied in high temperature superconducting (HTS) cuprates for their possible role in the pairing mechanism of these materials \cite{Scalapino1995329, orenstein,sidis-rev-res,RevModPhys-DopMottIns}. Although several studies have already been carried out by means of inelastic neutron scattering (INS)\cite{sidis-rev-res} on a number of cuprate compounds, the interpretation of the data remains highly controversial because of the lack of theoretical understanding of electronic correlations in realistic systems.

Recently, Cu $L_3$ edge resonant inelastic x-ray scattering (RIXS) \cite{ghiringhelli-prl-rixs-mag,Dean20153} has emerged as an alternative probe of the above excitations. This technique extends the energy range probed by INS to higher energies \cite{guarise-rixs-mag-prl} and also offers the advantage of measuring small single crystals. To the best of our knowledge, in HTS cuprates, RIXS has been hitherto employed to complete previous INS studies on well-known compounds. In the case of \LSCO{}, for example, the RIXS results found that magnetic excitations persist up to very high doping levels in regions of the Brillouin zone that are not easily probed by INS \cite{n-mat-dean-lascuo}.

The purpose of the present work is to study by means of RIXS the HTS cuprate parent compound \CCOC{} (CCOC), for which INS studies are infeasible because samples are only available as small, hygroscopic single crystals. This parent compound can be doped either  with sodium, \NaCCOC{} (Na-CCOC) \cite{hiroi,kohsaka-jacs}, or with vacancies, \VCCOC{} \cite{Yamada2005}. The motivation of our study is the simplicity of their single-layer tetragonal structure and the absence of structural instabilities that often jeopardize the study of more common cuprates, such as the aforementioned \LSCO{}. Moreover, the \CCOC{} system is the only HTS cuprate system composed exclusively of low $Z$ ions, with copper being the heaviest. This is an advantage for standard \textit{ab initio} density functional theory calculations, where large $Z$ ions pose problems for pseudopotential optimization. This feature is even more advantageous for advanced theoretical methods suitable to take into account correlation effects, such as quantum Monte Carlo, since they require one to treat accurately the spin-orbit coupling. In order to circumvent this difficulty, these quantum many-body calculations are mainly applied to systems with light atoms, where relativistic effects are negligible\cite{PhysRevX.4.031003,wagner2014,wagner-qmc-oxychlo}. Note that Ref.~\onlinecite{wagner-qmc-oxychlo} treats in particular \CCOC{}, although without reporting the values of the exchange $\mathbf{J}$. In this respect, \VCCOC{} and \NaCCOC{} are the most suitable example of such low-$Z$ systems among HTS cuprates. In addition, the superconducting compound \NaCCOC{} has already been studied by means of photoemission and scanning tunneling spectroscopy \cite{hiroi,k-shen,hanaguri}, therefore a RIXS study is expected to provide further insight into the electronic excitation spectrum. In the present work, by means of RIXS, we study the spin wave dispersion of \CCOC{}, the parent compound of the above HTS cuprate, and we extract the superexchange parameter $J$ using two different models.
                                                                                               
\begin{figure}[t]
\includegraphics[width=3.375in]{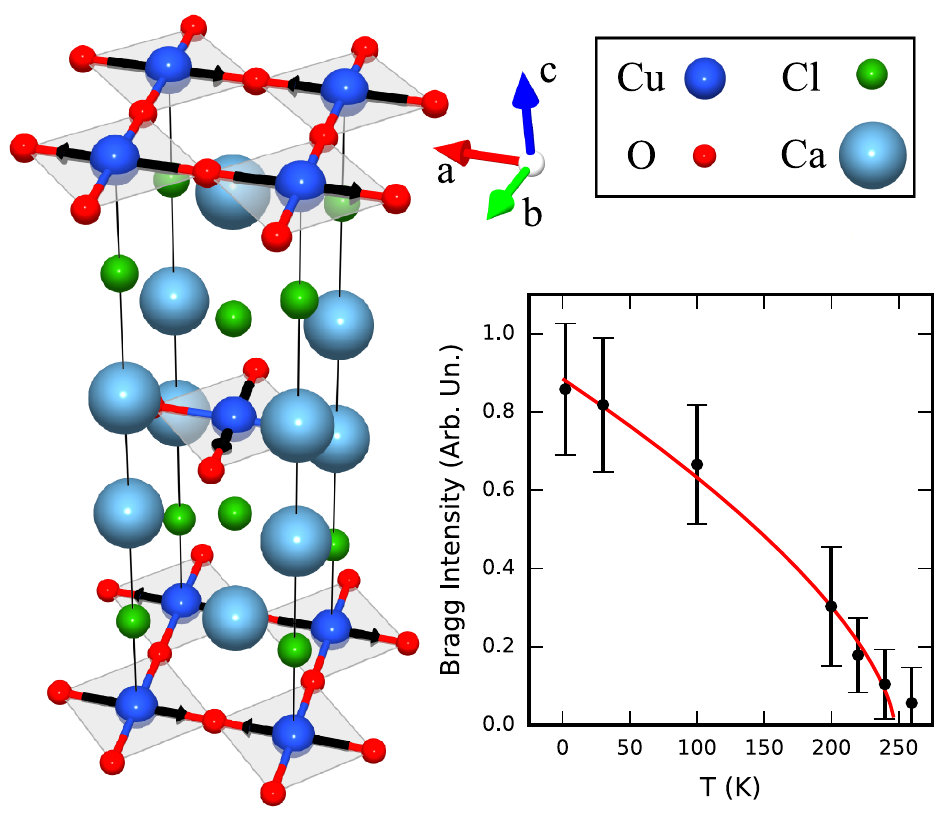}
	 \caption{\label{CCOC} (Color online) 
	 (top left) Tetragonal crystal structure\cite{VESTA} of \CCOC{} \cite{Yamada2005}. The square coordination of copper with its four nearest-neighbor oxygen ions in the CuO$_2$ planes is shown. The chlorine ions are located in the apical site above and below the copper. Black arrows indicate one of the possible magnetic structures consistent with neutron diffraction data\cite{vaknin_prb}.
     (bottom right) Temperature dependence of the fitted intensity of the averaged Bragg reflections ($\frac{1}{2}, \frac{1}{2}, \frac{5}{2}$)  and ($\frac{1}{2}, \frac{1}{2}, \frac{7}{2}$) and a power law fit (red).
	  }
\end{figure}

\section{\label{sec:methods} Experimental Methods}

\subsection{Crystal growth and characterization}

Single crystals of \CCOC{} were grown from CaCO$_3$, CuO, and CaCl$_2$ by solid state reaction, as described in detail elsewhere\cite{kohsaka-jacs,Yamada2005}. As shown in Fig.~\ref{CCOC}, \CCOC{} has a tetragonal K$_2$NiF$_4$-type structure (I4/mmm) \cite{ANIEBACK:ANIE197706741} with alternate stacking of (Ca,Cl)$_2$ and CuO$_2$ layers. The lattice parameters at ambient conditions are \textit{a}=\textit{b}=3.86735(2) \AA{} and \textit{c}=15.0412(1) \AA{}\cite{kohsaka-jacs,Yamada2005}. The crystals are easily cleaved along the \textit{ab}-plane due to the weak ionic bonds between adjacent layers. 

The single crystals of $\approx$2~mm width/height and $\approx$0.2~mm thickness were characterized using a commercial Bruker 4-circle kappa geometry diffractometer. A fixed Mo anode was used and the filtered K$_{\alpha}$ emission was collimated at 0.2 mm (3 mrad). A cryogenic N$_2$ flux was used to isolate the sample from humidity. The measurements yield unit-cell parameters in agreement with the literature \cite{kohsaka-jacs,Yamada2005} and also enabled us to determine the crystal orientation with respect to visible facets. The samples for RIXS measurements were subsequently glued on the holder with silver epoxy. Finally, ceramic posts were attached with the same epoxy in order to cleave the crystals in vacuum.

\CCOC{} is an antiferromagnetic insulator with a N\'eel temperature of T$_N$ = 247 $\pm$ 5 K \cite{vaknin_prb}. To check the magnetic state of the samples, we performed neutron scattering on the 1T spectrometer at Laboratoire Leon-Brillouin, using a sample from the same batch used for the RIXS experiment. We measured very weak magnetic reflections at low temperature for \textbf{q}=($\frac{1}{2}, \frac{1}{2}, \frac{\ell}{2}$) with $\ell$=2n+1 (n=0,...,4), but none for $\ell$=0, in agreement with Ref. \onlinecite{vaknin_prb}. The temperature dependence of the fitted Bragg intensity (average of the ($\frac{1}{2}, \frac{1}{2}, \frac{5}{2}$)  and ($\frac{1}{2}, \frac{1}{2}, \frac{7}{2}$) reflections) is shown in the bottom right of Fig.~\ref{CCOC} and a power law fit finds T$_N$ = 247 $\pm$ 6 K.

\begin{figure}[t]
\includegraphics[width=3.375in]{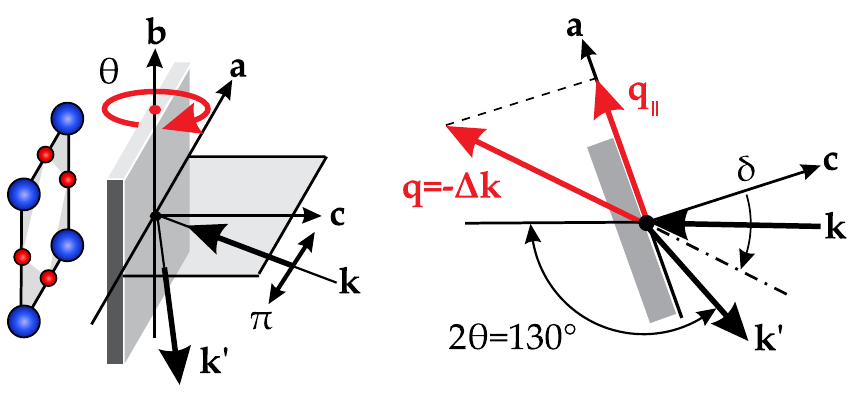}
	 \caption{\label{expgeo} (Color online) 
     RIXS geometry for measuring along $\langle$100$\rangle$ with $\pi$-polarization and grazing out emission (modified from Ref.~\onlinecite{MorettiSala2011}). 
The scattering angle 2$\theta$ is defined between the photon momentum of the incoming beam \textbf{k} and the direction where the analyzer collects the scattered beam \textbf{k'}. 2$\theta$ and the azimuthal angles are fixed, whereas the incident angle can be changed by a rotation, $\theta$, around the \textit{b}-axis. The incident angle defines $\delta$, which is the angle between the sample normal \textbf{c} and the transferred momentum \textbf{q} (red arrow), so that $\delta$ = 0 in specular reflection. The projection of \textbf{q} onto the sample's \textit{ab}-plane is denoted \textbf{q$_\parallel$}, which is 0 for $\delta$ = 0 and maximal for grazing geometries.  Measurements along $\langle$110$\rangle$ are done with the sample rotated 45$^{\circ}$ around the \textit{c}-axis.
	  }
\end{figure}

\subsection{\label{sec:RIXS}Resonant inelastic x-ray scattering}

RIXS measurements at the Cu L$_3$ edge (930 eV) were performed at the ADRESS beamline \cite{Strocov:bf5029,Schmitt201338} of the Swiss Light Source using the SAXES spectrometer \cite{saxes-rsi}. The samples were mounted in the ultra-vacuum manipulator cryostat of the experimental station. By applying a force on the aforementioned ceramic posts, the samples were cleaved \textit{in situ} under ultra-high vacuum and low temperature conditions to avoid hygroscopic damage of the cleaved surface. Their surface quality was confirmed by x-ray absorption spectroscopy. All spectra presented in this work were taken at 15~K. 

The experiment geometry is shown in Fig.~\ref{expgeo} and was similar to previous RIXS studies on cuprate parent compounds \cite{Dean20153}. We used $\pi$-polarized incident x-rays and a grazing exit geometry in order to enhance the single magnon spectral weight\cite{PhysRevLett.103.117003, PhysRevLett.105.167404, PhysRevB.85.064422, guarise-rixs-mag-prl, tacon-paramag, pol-dep-prb, dean-bisco-PRL}. The scattering angle was fixed at 2$\theta$ = 130$^{\circ}$, giving a constant momentum transfer to the sample of $q$ = 2\,k\,$\sin(\theta)$ = 0.85~\AA$^{-1}$. Although $q$ is fixed, its component in the \textit{ab}-plane, $q_\parallel$, can be changed by rotating the sample about the vertical axis (\textit{b}-axis in Fig.~\ref{expgeo}). For a given rotation, $\theta$, the deviation from specular reflection is given as $\delta = \theta_{specular} - \theta$, thus $q_\parallel$ = $q\sin(\delta)$. The minimum (maximum) $\delta$ used was +5$^{\circ}$ (+55$^{\circ}$) corresponding to $q_\parallel$ = +0.07~\AA$^{-1}$ ($q_\parallel$ = +0.70~\AA$^{-1}$). Therefore, in terms of reciprocal lattice units ($2\pi{}/a$) in the \textit{ab}-plane, we measured $\mathbf{q}_\parallel$ from (0.05,\,0) to (0.43,\,0) along $\langle$100$\rangle$ and from (0.03,\,0.03) to (0.3,\,0.3) along $\langle$110$\rangle$. In other terms (Fig.~\ref{disp} inset), we measured past the magnetic Brillouin zone along $\Gamma$-M, but well short of where thermal neutrons measure at M=(1/2,\,1/2). Along $\Gamma$-X we measured very close to the first Brillouin zone edge at X=(1/2,\,0).

\begin{figure}
\includegraphics[width=3.375in]{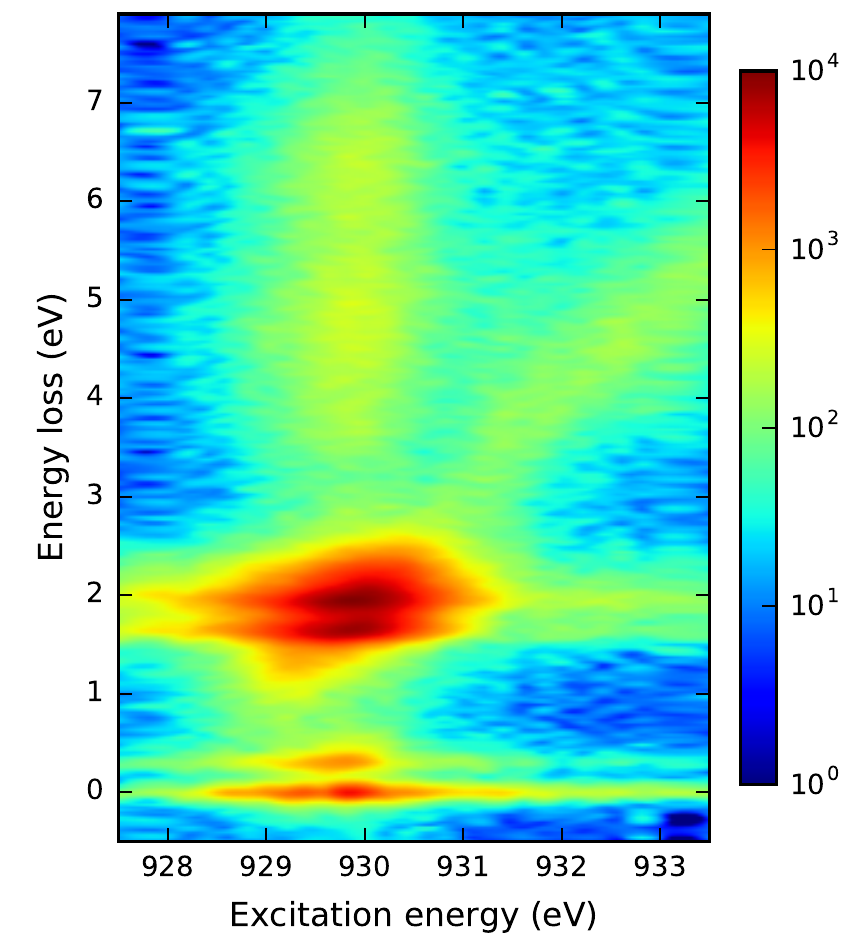}
	 \caption{\label{RIXS} (Color online) 
	 RIXS map at $\mathbf{q}_\parallel$ = (0.34,\,0) with $\pi$ incidence polarization showing the resonant behavior of the magnetic excitations, \textit{dd} excitations, and charge transfer excitations. Weak fluoresence is seen at high energy when the system is excited above the Cu L$_3$ edge threshold. The colormap is a logarithmic scale in arbitrary intensity units.
	  }
\end{figure}

\begin{figure}
\includegraphics[width=3.375in]{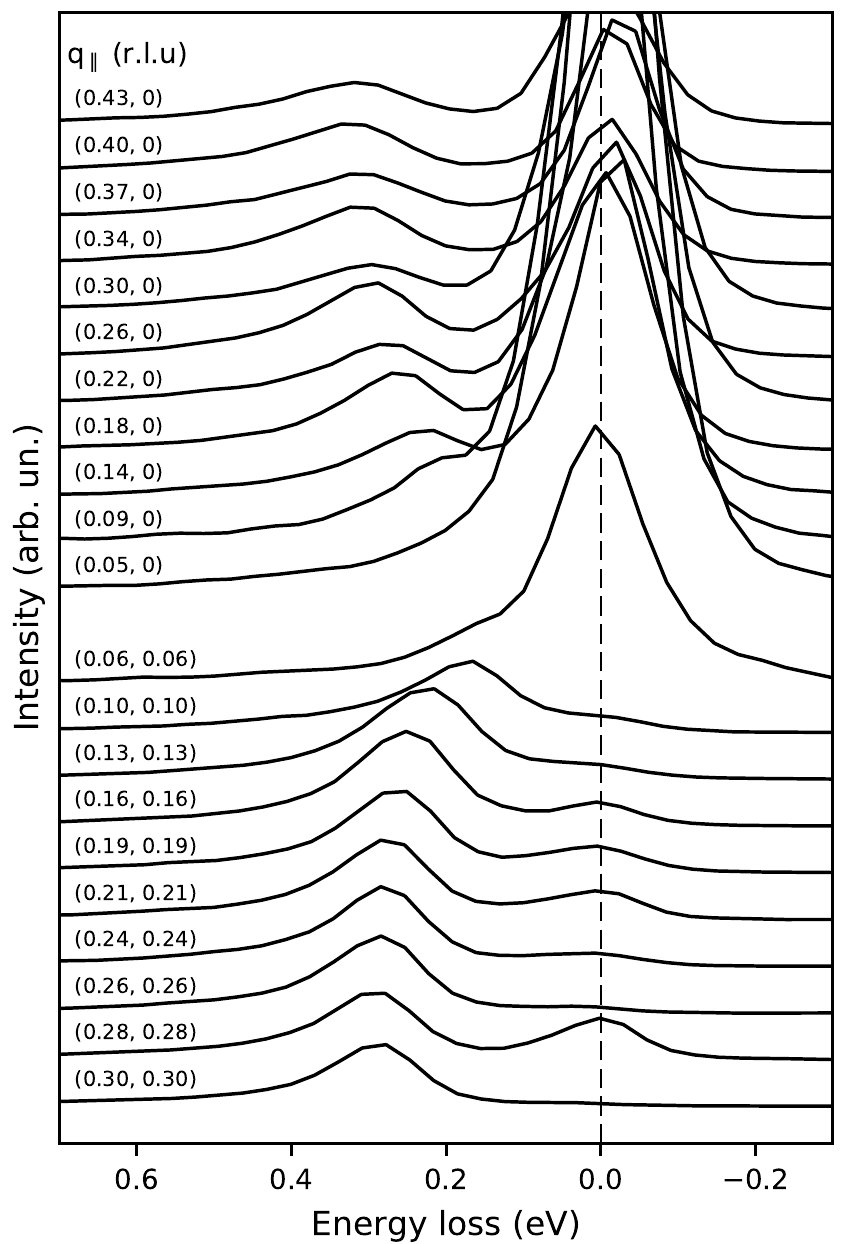}
	 \caption{\label{waterfall} 
    RIXS spectra showing the dispersion of the magnetic excitations along $\langle$100$\rangle$ (top) and $\langle$110$\rangle$ (bottom). Spectra are normalized by their \textit{dd} excitations.
	  }
\end{figure}

\begin{figure*}[t]
	\centering
	 \includegraphics{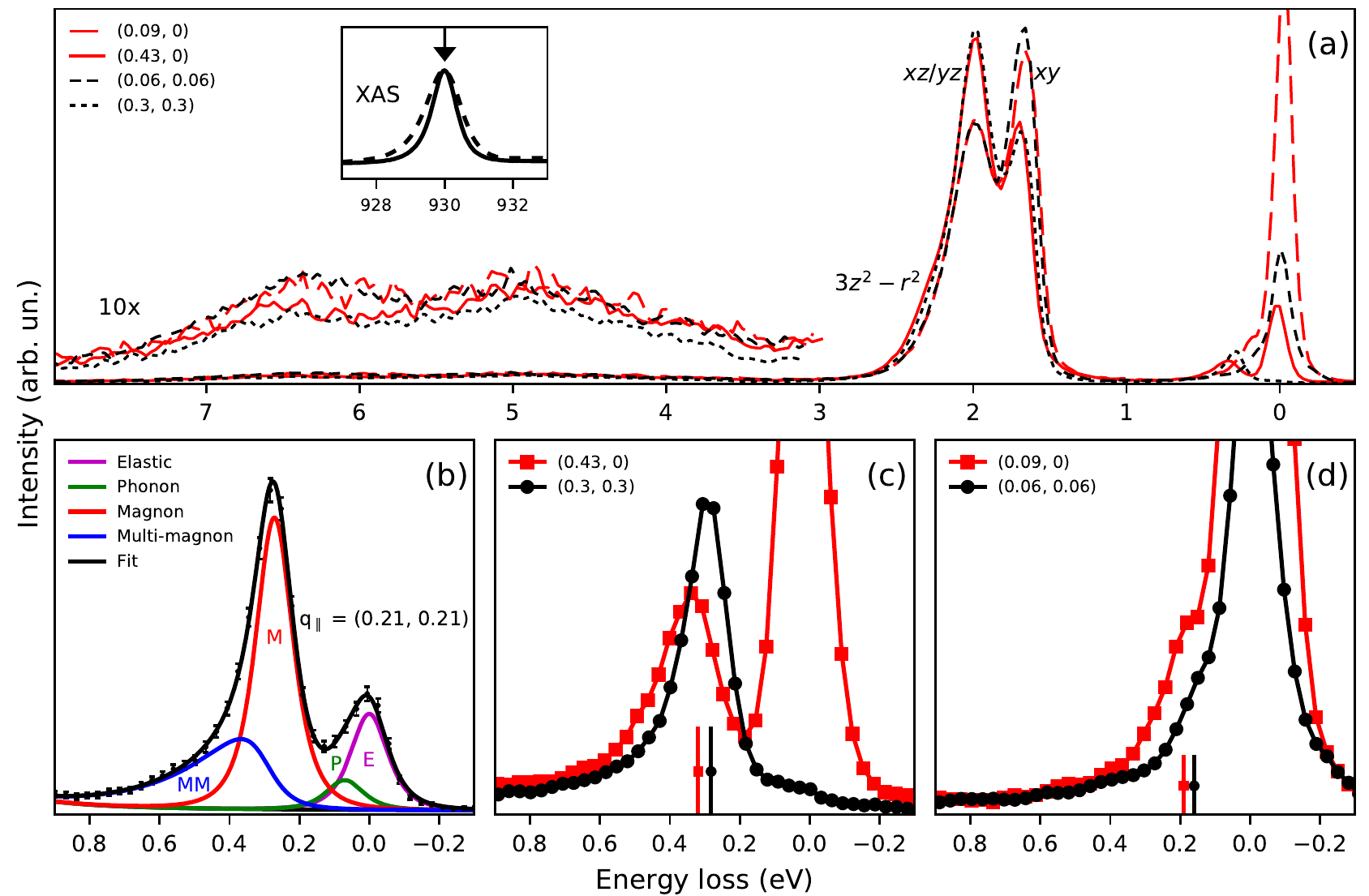}
	  \caption{\label{spectra} (Color online) Cu L$_3$ RIXS spectra of \CCOC{} at different in-plane transferred momentum, $\mathbf{q}_\parallel$, expressed as (\textit{h},\,\textit{k}) in reciprocal lattice units.
      (a) Representative RIXS spectra along $\langle$100$\rangle$ (red) and $\langle$110$\rangle$ (black). All spectra have been normalized to the area of the \textit{dd} excitations. The mid-infrared regions of these spectra are shown in (c,d), where vertical bars represent energy of single magnon found by fitting. The inset shows TEY-XAS (solid) and TFY-XAS (dashed), with an arrow indicating incident energy for our RIXS measurements. 
      (b) Example of fitting procedure at $\mathbf{q}_\parallel$ = (0.21,\,0.21) shown as black curve through data points. The elastic (E, magenta), phonon (P, green), and single magnon (M, red) peaks were resolution-limited and the multi-magnon (MM, blue) peak fitting is described in the text.
  }
\end{figure*}

\section{\label{sec:results} Results and discussion}

The RIXS map of \CCOC{} at $\mathbf{q}_\parallel$ = (0.34,\,0) shown in Fig.~\ref{RIXS} highlights the resonant behavior of the inelastic features. From lower to higher energy loss, one notes a mid-infrared peak between 0.1~eV and 0.6~eV, \textit{dd} excitations between 1~eV and 3~eV, and weak charge transfer excitations at higher energies. A weak fluoresence line is visible at energies above the Cu L$_3$ edge and intersects the \textit{dd} excitations at resonance. The spectral weight from this fluorescence line at resonance is unknown, but it is likely of the same order as the \textit{dd} excitations, as evidenced by the diagonal skew of the \textit{dd} excitations.

Fig.~\ref{waterfall} shows the RIXS spectra obtained along both directions focusing on the mid-infrared energy region, while Fig.~\ref{spectra}(a) shows the full energy region for $\delta$ = +10 and +55. The spectra are normalized to the area of the \textit{dd} excitations to account for the geometrical changes of the RIXS cross-section. There is an expected increase in elastic scattering near specular, i.e at (0.09,\,0) and (0.06,\,0.06). However, the elastic line for the sample aligned along $\langle$100$\rangle$ was large for all momentum transfers. These variations are likely due to finite surface quality after cleaving and did not impede accurate fitting.

\begin{figure}
	 \includegraphics[width=3.375in]{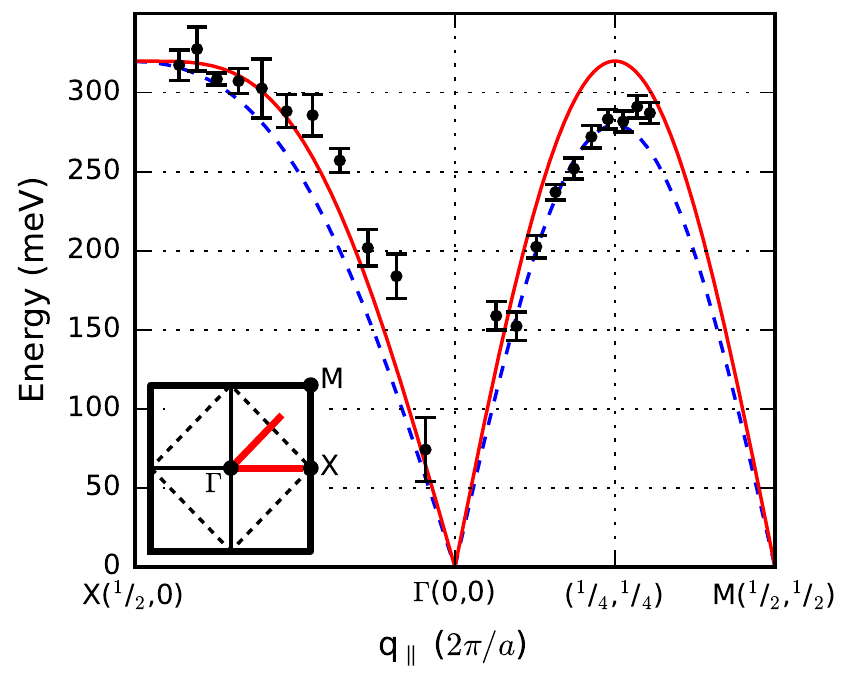}
	  \caption{\label{disp} (Color online) 
      	 Dispersion of \CCOC{} measured using Cu L$_3$ RIXS. The red, continuous line is a calculation for a classical spin-1/2 2D Heisenberg model with nearest-neighbor exchange and the blue, dashed line is a calculation including further exchange terms which is described in the text. (inset) 2D Brillouin zone showing high-symmetry points. The first Brillouin zone boundary is represented by a thick black square, while the magnetic Brillouin zone boundary is represented by a dashed line. The region where we measured is shown as two thick red lines along $\Gamma$-X and $\Gamma$-M. 
      }
\end{figure}

The mid-infrared feature is assigned as a magnon with a higher energy multi-magnon continuum. This assignment was done considering its dispersion (Fig.~\ref{waterfall},\ref{disp}) and past RIXS results on cuprate parent compounds in this experiment geometry \cite{guarise-rixs-mag-prl,Dean20153}. Furthermore, in our case, magnetic excitations are the only excitation in the mid-infrared energy region due to the $\approx$ 2~eV Mott gap. These spin excitations are the focus of our paper and are discussed below. 

The apical chlorine in \CCOC{} increases the tetragonal distortion much like for \SCOC{}, therefore based on Ref.~\onlinecite{MorettiSala2011} we assigned the \textit{dd} excitation at 1.70 eV to Cu-3d$_{xy}$, 1.99 eV to Cu-3d$_{xz/yz}$, and higher energies in the shoulder to Cu-3d$_{3z^2-r^2}$. The \textit{dd} excitations were not well fit following the technique of Ref.~\onlinecite{MorettiSala2011}, possibly due to fluorescence emission in this energy region or electron-phonon coupling\cite{PhysRevB.89.041104}. 

The broad charge transfer feature centered around 5.5 eV did not show dispersion or significant intensity variations, in agreement with Cu K edge RIXS\cite{Hasan2000a}. The author of Ref. \onlinecite{Hasan2000a} assigned this feature as transitions to an excited state composed of symmetric contributions of a central Cu-3d$_{x^2-y^2}$ orbital and the surrounding O-2p$_\sigma$ orbitals. Cu K edge RIXS also found a dispersive Mott excitation from 2.35 to 3.06 eV along $\Gamma$-X and from 2.34 eV to 4.14 eV along $\Gamma$-M. Therefore, the Mott excitation will fall under the \textit{dd} excitations for the majority of our momentum transfers, however, the Mott excitation at $\approx$~3.4 eV for $\mathbf{q}_\parallel$ = (0.3, 0.3) is not visible in our results (Fig.~\ref{spectra}(a)). 

A typical fit of the mid-infrared region is shown for $\mathbf{q}_\parallel$ = (0.21,\,0.21) in Fig.~\ref{spectra}(b) and the extracted magnon dispersion is shown in Fig.~\ref{disp}. The resolution function was measured on carbon tape and was well described by a Lorentzian squared function of 130~meV full-width at half-maximum. The elastic, phonon, and single magnon contributions were all resolution-limited. The multi-magnon excitation continuum was modeled as the resolution function convolved with a step function with subsequent exponential decay towards higher energy losses. The background was a Lorentzian tail of the form $y=A(x-x_0)^{-2} + c$. The energy of the phonon contribution is found around 60-70 meV with respect to the elastic, or $\sim$ 15-17 THz, roughly corresponding to the Debye cut-off frequency $\omega_D$ of \CCOC{} \cite{dastuto-ccocoPRB}. The major source of uncertainty for the magnon energy was determining the elastic energy, since the elastic line was irregular for the sample aligned along $\langle$100$\rangle$ and often too weak along $\langle$110$\rangle$. \textit{dd} excitations in undoped layered cuprates are known to be non-dispersive within current experimental accuracy\cite{MorettiSala2011}, therefore the elastic energy was fixed with respect to the Cu-3d$_{xz/yz}$ energy, which was found to be 1985 $\pm$ 5~meV from several spectra with well-defined elastic lines. 

The experimental and calculated dispersion along the two high-symmetry directions are shown together in Fig.~\ref{disp}. We use a classical $S$ = 1/2 2D Heisenberg model with higher order coupling to analyze our dispersion. The Hamiltonian is given by~\cite{coldea-prl-lco}:
\begin{eqnarray*}
\mathcal{H} = J \sum_{\langle i,j \rangle} \mathbf S_i \cdot \mathbf S_j + J' \sum_{\langle i,i' \rangle} \mathbf S_i \cdot \mathbf S_{i'} + J'' \sum_{\langle i,i'' \rangle} \mathbf S_i \cdot \mathbf S_{i''}
\\*
+ J_c \sum_{\langle i,j,k,l \rangle} \{ ( \mathbf S_i \cdot \mathbf S_j ) ( \mathbf S_k \cdot \mathbf S_l ) + ( \mathbf S_i \cdot \mathbf S_l ) ( \mathbf S_k \cdot \mathbf S_j ) 
\\*
- ( \mathbf S_i \cdot \mathbf S_k ) ( \mathbf S_j \cdot \mathbf S_l ) \}
\end{eqnarray*}
where we include first-, second-, and third-nearest neighbor exchange terms, as well as a ring exchange term ($J$, $J'$, $J''$, and $J_c$). Within classic linear spin-wave theory \cite{PhysRevB.45.7889,spinwave-code} this leads to a dispersion relation given by\cite{coldea-prl-lco} $\hbar\omega_\mathbf{q}=2 Z_C(\mathbf q) \sqrt{A_{\mathbf q}^2 - B_{\mathbf q}^2}$ where $A_{\mathbf q}^2 = J - J_c/2 - (J' - J_c/4)(1-\nu_h \nu_k) - J''[1 - (\nu_{2h} + \nu_{2k})/2]$, $B_{\mathbf q}^2 = (J - J_c/2)(\nu_h + \nu_k)/2$, $\nu_x = cos(2 \pi x)$, and $Z_C(\mathbf q)$ is a spin renormalization factor\cite{spin05quant-fluct,coldea-prl-lco}. 

As a first approximation we consider only the first term in the Hamiltonian, which corresponds to only nearest-neighbor exchange. In this isotropic case the dispersion relation above reduces to $\hbar\omega_\mathbf{q}=2 J Z_C \sqrt{1-\lbrack\cos(2\pi h) +\cos(2\pi k)\rbrack^2/4}$, where $Z_c$ = 1.18 is a constant\cite{spin05quant-fluct}. The calculation for our data is shown in Fig.~\ref{disp} as a solid red line, obtained both analytically and using the ``SPINWAVE'' code \cite{spinwave-code}, as a check. The energy at the zone boundary peaks at $2JZ_C$ = 320 $\pm$ 10~meV, which gives $J$ = 135 $\pm$ 4~meV. For \LCO{} and \SCOC{}, the zone boundary energy is 314 $\pm$ 7~meV and 310~meV respectively, which corresponds to $J$ = 133 $\pm$ 3~meV and $J$ = 131~meV respectively\cite{coldea-prl-lco,guarise-rixs-mag-prl}.

Note the 40 $\pm$ 10~meV energy difference along the magnetic Brillouin zone boundary (MBZB) between X and M. This MBZB dispersion is an indication of non-negligible magnetic interactions beyond nearest-neighbors\cite{guarise-rixs-mag-prl, coldea-prl-lco,dalla_piazza_rapidB}. Following Ref.~\onlinecite{coldea-prl-lco}, we parametrize the above Hamiltonian with a single band Hubbard model with U, the on-site repulsion, and t, the nearest-neighbor hopping. Expanding the Hubbard Hamiltonian to order t$^4$, we find $J = 4t^2/U - 24t^4/U^3$, $J_c = 80t^4/U^3$, and $J' = J'' = 4t^4/U^3$. We assume the spin renormalization is constant, $Z_c(\mathbf q) \approx Z_c$, which introduces an error less than the uncertainty of our data\cite{coldea-prl-lco}. Within this model, it can be shown \cite{2016arXiv160905405P} that the maximum energy at X is given by $E_{max}  = 2Z_C(J - J_c/10)$ and the energy dispersion along the MBZB is given as $\Delta E_{MBZB} = 3 Z_C J_c/5$. We can use our experimental dispersion to fix $E_{max}$ = 320~meV and $\Delta E_{MBZB}$ = 40~meV, which uniquely determines U = 2.2~eV and t = 295~meV. The corresponding superexchange parameter is $J$ = 141~meV, versus $J$ = 146~meV for \LCO{} and $J$ = 144~meV for \SCOC{}. The calculated dispersion using these values is shown in Fig.~\ref{disp} as a dashed blue line. The MBZB dispersion is well fit, however the energy along $\langle$100$\rangle$ is underestimated, indicating the need to include further hopping terms in the Hubbard model\cite{PhysRevB.85.100508, PhysRevB.79.235130}. Furthermore, our values of U and t are unphysical, even if they are similar to those found in \LCO{} at 10~K using this approach\cite{coldea-prl-lco} (U = 2.2~eV and t = 300~meV). They are in disagreement with photoemission results\cite{Ronning2067} and U = 7.5t is less than the tight binding bandwidth\cite{PhysRevB.79.235130} of 8t. Inclusion of further hopping terms is beyond the scope of this paper, however they will not fundamentally change the determination of the superexchange parameter $J$.

The fact that all three cuprates discussed above have a very similar $E_{max}$ is a bit surprising. The simplistic scaling relation\cite{harrison1980electronic} $J \propto {d_{NN}}^4$ based on the intra-planar Cu NN distance would predict a 7\% softening of \CCOC{} with respect to \LCO{} ($d_{NN}$ = 3.803 \AA{})\cite{PhysRevB.41.1926} and an 11\% hardening with respect to \SCOC{} ($d_{NN}$ = 3.975 \AA{})\cite{PhysRevB.41.1926}. 

On the other hand, these three cuprates have different $\Delta E_{MBZB}$, with \LCO{} being smaller (22 $\pm$ 10~meV) and \SCOC{} being larger (70~meV). With further exchange terms\cite{ivashko-damped-sw-lsco} it is found that the dispersion scales as $(t'/t)^2$, where $t'$ is the next-nearest-neighbor hopping. This second hopping term is typically decreased due to apical hybridization\cite{PhysRevLett.87.047003}, therefore we would expect greater dispersion for longer apical bonds lengths. This is indeed the trend we see for these three compounds: \SCOC{} (2.8612~\AA{}) $>$ \CCOC{} (2.734~\AA{}) $>$ \LCO{} (2.416~\AA{}). If this interpretation is correct, then our assignment of the shoulder in the \textit{dd} excitations to Cu-3d$_{3z^2-r^2}$ is likely incorrect since we would then expect E$_{3z^2-r^2}$ for \CCOC{} to be less than 1.97~eV (\SCOC{}) and more than 1.7 eV (\LCO{})\cite{MorettiSala2011}.

Although Ref.~\onlinecite{wagner-qmc-oxychlo} did not calculate $J$, the current uncertainty in QMC calculations allows a rough comparison between them and experiment. QMC calculations\cite{wagner2014,PhysRevX.4.031003} have found $J$ = 160(13)~meV for \LCO{}, $J$ = 140(20)~meV for CaCuO$_2$, and $J$ = 159(14)~meV for Ca$_2$CuO$_3$. The value found for \LCO{} is quite different from its experimental value, possibly due to relativistic effects in the La atoms. CaCuO$_2$ and \CCOC{} are both composed of CuO$_{2}$ planes with interplanar Ca atoms, however CaCuO$_2$ lacks any apical ligand. Nonetheless, its calculated value matches quite well our results above, much better than the Cu chain system of Ca$_2$CuO$_3$ which has apical oxygens, emphasizing the important role that the apical ligands play in intraplanar(chain) exchange.

\section{Conclusions}
In conclusion, the present Cu L$_3$ edge RIXS study enabled us to determine the spin wave dispersion along the two high-symmetry directions of \CCOC{}, an undoped antiferromagnetic HTS cuprate parent compound containing only low-Z elements. In first approximation, the data are explained within a simple S = 1/2 2D Heisenberg model with a nearest-neighbor exchange term $J$ = 135 $\pm$ 4~meV, taking into account spin quantum fluctuation renormalization. Including next-nearest-neighbor contributions, our estimate is increased to $J$ = 141~meV. To the best of our knowledge, this is the first measurement of the spin-wave dispersion and of its zone-boundary energy in \CCOC{}, noting that INS experiments are currently infeasible and two-magnon Raman scattering has not been performed yet. We believe that the present low-$Z$ cuprate \CCOC{} is an ideal playground for future quantum many-body theoretical models of HTS cuprates. Our RIXS results combined with the future results of these models will offer a unique comparison between experiment and state-of-the-art theory of correlated electron systems.

\begin{acknowledgments}
The authors acknowledge the Paul Scherrer Institut, Villigen-PSI, Switzerland for provision of synchrotron radiation beamtime at beamline X03MA, "ADRESS" of the Swiss Light Source, as well as LLB and KIT for providing neutron beamtime on the 1T spectrometer, and would like to thank Yvan Sidis for his assistance. They are grateful to Jean-Pascal Rueff, Sylvain Petit, and  Marco Moretti for fruitful discussions, as well as Lise-Marie Chamoreau for her assistance in sample preparation. 
We are very grateful to Sylvain Petit for his help with ``SPINWAVE'' code \cite{spinwave-code}.
B.L acknowledges financial support from the French state funds managed by the ANR within the ``Investissements d'Avenir'' programme under reference  ANR-11-IDEX-0004-02, and within the framework of the Cluster of Excellence MATISSE led by Sorbonne Universit\'{e} and from the LLB/SOLEIL PhD fellowship program. This material is based upon work supported by the U.S. Department of Energy, Office of Basic Energy Sciences, Early Career Award Program under Award Number 1047478. Brookhaven National Laboratory was supported by the U.S. Department of Energy, Office of Science, Office of Basic Energy Sciences, under Contract No. DE-SC00112704. J.P. and T.S. acknowledge financial support through the Dysenos AG by Kabelwerke Brugg AG Holding, Fachhochschule Nordwestschweiz, and the Paul Scherrer Institut. M.D. acknowledges financial support from the Swiss National Science Foundation within the D-A-CH programme (SNSF Research Grant 200021L 141325). M.d'A. acknowledges travel funding from the E.C. under the 7th Framework Program within the CALIPSO Transnational Access support. This work was written on the collaborative OVERLEAF platform \cite{overleaf}.
\end{acknowledgments}

\bibliography{ccoc_rixs_sw}
\bibliographystyle{apsrev}

\end{document}